\documentclass[10pt,a4paper]{jpconf}
\usepackage{graphicx}
\usepackage{color}
\begin{document}
\title{Loop quantum effect and the fate of tachyon field collapse}

\author{Y. Tavakoli~$^{1,*}$, P. Vargas Moniz~$^{1,2,\dag}$, J. Marto~$^{1,\ddag}$, A. H. Ziaie~$^{3,\S}$}
\address{\small $^{1}$ \emph{Departamento de F\'{\i}sica, Universidade da Beira Interior, 6200 Covilh\~{a}, Portugal}}
\address{\small $^{2}$ \emph{CENTRA, IST, Av. Rovisco Pais, 1, Lisboa, Portugal}}
\address{\small $^{3}$ \emph{Department of Physics, Shahid Beheshti University, Evin, Tehran 19839, Iran}\\
\small E-mail:~\mbox{$^{*}~$tavakoli@ubi.pt}, \mbox{$^{\dag}~$pmoniz@ubi.pt}, \mbox{$^{\ddag}~$jmarto@ubi.pt}, \mbox{$^{\S}~$ah.ziaie@gmail.com}}





\begin{abstract}
We study the fate of gravitational collapse of a tachyon field matter. In presence of an inverse square potential a  black hole forms. Loop quantum corrections  lead to the avoidance of classical singularities, which is followed by an outward flux of energy.
\end{abstract}


\section{Gravitational collapse of a tachyon matter}
\medskip
We discuss the spacetime region inside the collapsing sphere which will
contain the chosen matter content. In this region, an homogeneous and isotropic Friedmann-
Robertson-Walker (FRW) metric will be considered as the setting for the gravitational
collapse in comoving coordinates. However, in order to study the whole spacetime, we
must match this interior spacetime to a suitable exterior region. In the model herein, it
is convenient to consider a spherically symmetric and inhomogeneous spacetime such as the
Schwarzschild or the generalized Vaidya geometries to model the spacetime outside
the collapsing sphere  (cf. see \cite{YPJH} for a discussion on the whole spacetime structure including the exterior geometry).
Let us nevertheless be more precise for the benefit of
the reader: There is a particular setting which has been of recent use, namely in investigations
of loop effects in gravitational collapse involving a scalar field; it is the marginally
bound case ($k$ = 0), where the interior spacetime is dynamical and is parameterized by a
line element as follows,
\begin{equation}
ds^{2}=-dt^{2}+a^{2}(t)[dr^{2}+r^{2}d\Omega^{2}],
\label{metric}
\end{equation}
where $t$ is the proper time for a falling observer whose geodesic trajectories are labeled by
the comoving radial coordinate $r$, and $d\Omega^2$
being the standard line element on the unit two-
sphere; we have set the units $8\pi G = c = 1$. The dynamical evolution from Einstein's field
equations for this marginal bound case regarding the interior region, can be presented as \cite{2}
\begin{equation}
\rho=\frac{F_{,r}}{R^2R_{,r}}, ~~~~p=-\frac{\dot{F}}{R^2\dot{R}},~~~~\dot{R}^2=\frac{F}{R},
\label{einstein}
\end{equation}
where $t$ and $r$ are the proper time and the comoving radial coordinate for a falling observer, respectively;
we have set the units $8\pi G=c=1$. 
The quantity $F(R)$  is the mass function with $R(t,r)=ra(t)$ being the area radius of the collapsing
shell. Furthermore, $\rho$ and $p$ are energy density and pressure of the collapsing matter, respectively. The notation ``$,r"$ and ``$\cdot$" denote a derivative with respect
to $r$ and $t$, respectively.

Integration of Eq. (\ref{einstein}) gives the
following relation for the mass function as
$F=\frac{1}{3}\rho(t)R^3$. A spherically symmetric spacetime containing a trapped region is characterized  by the corresponding
boundary (trapped) surface, extracted from the equation $F=R$. In
the region where the mass function satisfies the relation $F>R$, it
describes a trapped region and the regions in which the mass
function is less than the area radius i.e., $F<R$, are not trapped.

To investigate the gravitational collapse, we will consider herein a homogeneous tachyon field $\phi(t)$, as matter content of the collapsing cloud. Employing A. Sen's effective action,
the corresponding energy density and pressure for tachyon field are given by \cite{3}
\begin{eqnarray}
\rho(\phi)=3H^2=\frac{V(\phi)}{\sqrt{1-\dot{\phi}^{2}}},~~~~ p(\phi)=-V(\phi)\sqrt{1-\dot{\phi}^{2}},
\label{energy}
\end{eqnarray}
where $V(\phi)$ is the tachyon potential and is assumed to be of the form $V(\phi)=V_0\phi^{-2}$. Using the equation for the
energy conservation, we get the field equation for $\phi$ as
\begin{equation}
\ddot{\phi}=-(1-\dot{\phi}^2)\left[3H\dot{\phi}+\frac{V_{,\phi}}{V}\right],
\label{field}
\end{equation}
where ``$,\phi$" denotes the derivative  with respect to the tachyon
field and $H=\dot{a}/a$ is the Hubble rate which for collapsing scenario is $H<0$. Using Eq. (\ref{energy}) the equation of state becomes
$w=p(t)/\rho(t)=\dot{\phi}^2-1$.

Let us study the collapse of a tachyon field from a dynamical
system perspective. We  define  a new time variable $\tau$ (instead of the proper time
$t$) for our  collapsing system as
$\tau \equiv -\log (a/a_0)^{3}$, which is  defined in the interval $0<\tau<\infty$.
For any time dependent function $f=f(t)$ we can
write $df/d\tau\equiv -\dot{f}/3H$, where a dot over a variable denotes its  derivative  with respect to
time $t$.

Introducing   the  new set of dynamical variables
$x\equiv\dot{\phi}$ and $y\equiv V/3H^2$, the Friedmann constraint equation in (\ref{energy}) becomes
$y/\sqrt{1-x^2}=1$. From Eqs. (\ref{energy}) and (\ref{field}) the corresponding  autonomous system then reads
\begin{equation}
\frac{dx}{d\tau}=\left(\frac{\lambda}{\sqrt{3}}\sqrt{y}+x\right)(1-x^2),~~~~\frac{dy}{d\tau}=-xy\left(\frac{\lambda}{\sqrt{3}}\sqrt{y}+x\right),
\label{DyClX}
\end{equation}
where
$\lambda \equiv -V_{,\phi}/V^{3/2}=2/\sqrt{V_0}=constant$.  The critical points
$(x_c, y_c)$ of the autonomous system can be obtained easily as: $(a)$ at $ (1, 0)$, $(b)$ at
$(-1, 0)$, and $(c)$ at $(-\lambda\sqrt{y_0/3}, y_0)$, where
$y_0=-\lambda^2/6+\sqrt{(\lambda^2/6)^2+1}>0$.
From the eigenvalues of the Jacobi matrix, the stability of the system can be
determined;
\begin{itemize}
  \item Fixed points $(a)$ and $(b)$  are asymptotically \emph{stable}
nodes (attractor) allowing to select a class of solutions  in which, asymptotic  time
dependence of the tachyon field   can be estimated as
$\phi(t) \simeq  \pm t+\phi_0$, where `$+$' and `$-$' denotes fixed points $(a)$ and $(b)$, respectively.
From Eq. (\ref{energy}) it is seen that, the effective pressure asymptotically vanishes.
This suggests a dust-like behavior where $p\simeq 0$ \cite{homodust}.
Then, the energy density of collapse can be obtained as $\rho=\rho_{\phi}\propto
1/a^3$; this expression indicates, therefore  that, $\rho_{\phi}$
diverges as $a\rightarrow0$.
  \item The critical point $(c)$ is  asymptotically \emph{unstable} (source); it can not be considered as a physically relevant solution for the collapse, and hence we ignore its studying here.
\end{itemize}
The occurrence of spacetime singularities is proved
by the existence of incomplete future or past directed
non-spacelike geodesics. Then, it is required that the Kretschmann invariant $K=R_{abcd}R^{abcd}=12[\left(\ddot{a}/a\right)^2+\left(\dot{a}/a\right)^4]$ gets
unboundedly large values as the singularity is approached along
the non-spacelike geodesics terminating there. If such a condition
is satisfied, then the physical system should be considered to
exhibit a curvature singularity \cite{2}. Therein one can observe that, this quantity diverges as the
physical area radius vanishes, consequently signaling the occurrence of a
curvature singularity; the collapse ends up with a black hole formation \cite{YPJH}.

\section{Semiclassical tachyon field collapse}
\medskip
In this section we will study the loop quantum effects (of the inverse triad type) on tachyon field collapse. To investigate the
fate of the collapse we will consider the modified dynamical equation of the system in the semiclassical limit.
The modified Friedmann and the Raychaudhury equations for the  tachyon
field in the semiclassical regime $a_i<a\ll a_*$ are given, respectively by \cite{3}
\begin{eqnarray}
\rho_{\phi}^{loop}=3H^2=\frac{V(\phi)}{\sqrt{1-A^{-1}q^{-15}\dot{\phi}^2}},\ \ \ \ \ \ p_{\phi}^{loop}=2\dot{H}+3H^2=-\left(1+4\frac{\dot{\phi}^2}{Aq^{15}}\right)\rho_{\phi}^{loop}.
\label{energyloop}
\end{eqnarray}
Then, using the energy conservation equation, the modified equation of motion in the semiclassical limit becomes
\begin{equation}
\ddot{\phi}-12H\dot{\phi}\left(\frac{7}{2}-\frac{\dot{\phi}^2}{Aq^{15}}\right)+
\left(Aq^{15}-\dot{\phi}^2\right)\frac{V_{,\phi}}{V}=0,
\label{loopField}
\end{equation}
where $A\equiv (12/7)^{12}$. Also $a_{*}=\sqrt{j/3}a_i$ is a critical scale at which the eigenvalue of the inverse scale factor has a
power-law dependance on the scale factor; $j$ is the half-integer free quantization parameter, $a_{i}\equiv \sqrt{\gamma}\ell_{{\footnotesize\textrm{P}}}$ is the scale above which a classical continuous spacetime can be defined and below which the spacetime is discrete, and $\ell_{{\footnotesize\textrm{P}}}$ is the Planck length \cite{Bojowald1}. The modified mass function within this regime ($a\ll a_*$) takes the form $F^{loop}=1/3\rho_{\phi}^{loop}R^3$.
The behavior of this effective mass function determines whether trapped surfaces will form during  the collapse procedure, within
the semiclassical regime.

To study  the collapsing model using a
dynamical system description
we introduce a new time variable $\sigma$ given by
$\sigma\equiv -\log
q^{3/2}$, where $q=(a/a_*)^2$,   being defined in the interval $0<\sigma<\infty$;
the limit $\sigma\rightarrow 0$ corresponds to the initial condition of
the collapsing system ($a\rightarrow a_*$) and the limit
$\sigma\rightarrow \infty$ corresponds to $a\rightarrow 0$. For any time
dependent function $f$, we write
$df/d\sigma\equiv -\dot{f}/3H$,
where dot denotes the derivative  with respect to $t$.

We herein make use of  a \emph{new} set of convenient dynamical
variables: $\tilde{x}\equiv A^{-1/2}q^{-15/2}\dot{\phi},\,\,\, \tilde{y}\equiv V/3H^2,\,\,\, \tilde{z}\equiv
A^{1/2}q^{15/2}$.
Then,  Eqs. (\ref{energyloop})-(\ref{loopField}) in terms of  $\tilde{x}$, $\tilde{y}$, and $\tilde{z}$ are presented as
\begin{eqnarray}
\frac{d\tilde{x}}{d\sigma}=-\frac{\xi}{\sqrt{3}} \tilde{z}\sqrt{\tilde{y}}\left(1-\tilde{x}^2\right)+\tilde{x}\left(4\tilde{x}^2-9\right), \ \ \ \ \ \ \ \nonumber \\
\frac{d\tilde{y}}{d\sigma}=-\frac{\xi}{\sqrt{3}} \tilde{z}\sqrt{\tilde{y}}\left(1-\tilde{x}^2\right)+\tilde{x}\left(4\tilde{x}^2-9\right),\ \ \ \ \ \ \ \frac{d\tilde{z}}{d\sigma}=-5\tilde{z},
\label{dynZ}
\end{eqnarray}
where $V(\phi)=\beta\phi^{-2}$, which  brings  $\xi\equiv -V_{,\phi}/V^{3/2}=2/\sqrt{\beta}$ as a constant. The Hamiltonian constraint equation (\ref{energyloop}) in terms of new variables takes the form
$\tilde{y}/\sqrt{1-\tilde{x}^2}=1$.
The properties of each critical point will be determined by the eigenvalues of the
$(3\times3)$-Jacobi matrix.
Finally, for the dynamical system (\ref{dynZ}), there is, under the constraints,  just one physical fixed
point for the physical system: $(p_o)$ as $ (0, 1, 0)$.
In order to discuss the stability of $(p_o)$,
we have to determine the eigenvalues (and eigenvectors) of then Jacobi matrix. This analysis shows that,
all eigenvalues are real but one   is  zero,  the rest being
negative implying  that, the autonomous system is nonlinear
with a non-hyperbolic point.   The  asymptotic
properties cannot be simply determined by  linearization  and we
need to resort to other method: the center manifold theorem.
Using the center manifold analysis one can show that, $(p_o)$ is an asymptotically \emph{stable} critical point (for a more detailed and complete discussion on the above elements, the reader may consult ref. \cite{YPJH2}).

The solution for the effective energy density, the effective pressure 
in the neighborhood of
the critical point $(p_o)$
can be easily obtained that
\begin{eqnarray}
\rho_{\phi}^{loop}\simeq V(\phi)\left(1+\frac{1}{2}\frac{\dot{\phi}^2}{Aq^{15}}\right),~~~~p_{\phi}^{loop}\simeq -V(\phi)\left(1+\frac{9}{2}\frac{\dot{\phi}^2}{Aq^{15}}\right).
\label{energydyn}
\end{eqnarray}
For a sufficient choice of initial condition, this fixed point leads to a solution for the scale factor of the collapsing system as $a^{42}(\phi)=42\sqrt{\beta/3}\ln\left|\phi_s/\phi\right|$, where $\phi_s$ is a constant of integration.
Then, it is seen that, when the tachyon field approaches the finite value as $\phi\rightarrow
\phi_s$, the scale factor vanishes as $a\rightarrow 0$. On the other hand, the
potential of tachyon field decreases from its initial value, and
approaches a finite value at $a=0$. Furthermore, the effective
energy density which is given by $\rho_{\phi}^{loop}\approx V(\phi)=
\beta/\phi^2$ does not blow up, becoming instead  small and
remaining finite. On the other hand, the effective mass function is
$F^{loop}/R\approx (\beta/3\phi^2)r^2a^2$ and vanishes as
$a\rightarrow 0$; that is, no trapped surfaces form for collapsing
shells. Moreover, the effective pressure in Eq. (\ref{energydyn}) evolves such that
$p_{\phi}^{loop}\approx -\rho_{\phi}^{loop}=-\beta/\phi^2$,
remaining  finite and negative, giving rise to an outward flux of
energy in semiclassical regime.

Semiclassical analysis for the Kretschmann scalar shows that, it reaches a local maximum as the singularity is approached and then converges
to finite values. Consequently this behavior can be interpreted as an effective singularity avoidance.

\section{Conclusion}
\medskip

In this paper we considered the marginally bound FRW
as a spherically symmetric model for an homogeneous interior spacetime of a gravitational
collapsing system constituted by a tachyon matter field with the potential of an inverse square
form. We investigated the classical and
semiclassical regimes which the latter characterized by inverse
triad corrections in order to establish the corresponding
final state. Within the classical setting, we
found analytically a situation where the tachyon evolved
towards a dust-like asymptotic behavior; this suggested
a black hole forming. In semiclassical regime, loop quantum corrections (an inverse triad type), pointed to the regularity of the spacetime geometry
and the energy density of the system, where classical trapped surfaces were absent therein.
These conditions indicated that, the spacetime in
semiclassical collapse is singularity-free, hence there is
neither a naked singularity nor a black hole forming as final state of collapse. In other words, loop quantum
corrections of the inverse triad type seem to induce
an outward flux of energy at the final state of the collapse.

\section{Acknowledgments}
\medskip
YT is supported by the Portuguese
Agency Funda\c{c}\~{a}o para a Ci\^{e}ncia e Tecnologia (FCT)
through the
fellowship SFRH/BD/43709/2008. He also would like to thank M. Bojowald for the useful discussion. This research work was also supported by the grant CERN/FP/109351/2009
and CERN/FP/116373/2010.


\section{References}
\medskip
\begin{thebibliography}{9}

\bibitem{YPJH} Y. Tavakoli, P. Vargas Moniz, J. Marto, and A. H. Ziaie, \emph{Gravitational collapse with tachyon field and barotropic fluid} [arXiv: gr-qc/0911.1428].

\bibitem{2} P. Joshi, {\it Gravitational Collapse and Space-Time Singularities}, (Cambridge University Press, 2007).

\bibitem{3} A. A. Sen, Phys. Rev. D \textbf{74}, 043501 (2006).

\bibitem{homodust} J. R. Oppenheimer, and H. Snyder, Phys. Rev. \textbf{56}, 455 (1939).

\bibitem{Bojowald1} M. Bojowald, Class. Quant. Grav. \textbf{19}, 5113 (2002).

\bibitem{YPJH2} Y. Tavakoli, P. Vargas Moniz, J. Marto, and A. H. Ziaie, \emph{Semiclassical collapse with tachyon field and barotropic fluid} [in preparation].

\end{thebibliography}
\end{document}